\documentclass[preprint]{aastex}

\newcommand{\cii}{C~II $\lambda$1335}

\newcommand{\siiv}{Si~IV $\lambda$1400}
\newcommand{\siii}{Si~II $\lambda$1526}
\newcommand{\civ}{C~IV $\lambda$1550}

\newcommand{\Ms}{M$_{\odot}$}
\newcommand{\Mup}{$M_{\rm{up}}$}
\newcommand{\Mlow}{$M_{\rm{low}}$}

\newcommand{\Zs}{Z$_{\odot}$}
\newcommand{\kms}{km~s$^{-1}$}

\tighten
\journalid{}{}
\articleid{}{}

\begin{document}

\title{The stellar content of the super star clusters in NGC~1569$^{\star}$
\footnotetext{$^{\star}$Based on observations with the 
NASA/ESA Hubble Space Telescope, obtained at the Space Telescope Science 
Institute, which is operated by AURA for NASA under contract NAS5-26555}
}

\author{Livia Origlia$^1$, Claus Leitherer$^2$, Alessandra Aloisi$^{2}$,
        Laura Greggio$^{1,3}$, and Monica Tosi$^1$}

\vspace{0.25in}

\affil{$^1$ Osservatorio Astronomico di Bologna, Via Ranzani 1,
       I-40127 Bologna, Italy\\
       e-mail: origlia@bo.astro.it, greggio@bo.astro.it,
tosi@bo.astro.it}

\affil{$^2$ Space Telescope Science Institute,
       3700 San Martin Drive, Baltimore, MD 21218\\
       e-mail: leitherer@stsci.edu, aloisi@stsci.edu}
     
\affil{$^3$ Universitaets Sternwarte Muenchen,
       Scheinerstrasse 1, D-81679 Muenchen, Germany\\
       e-mail: greggio@usm.uni-muenchen.de}

\begin{abstract}
We discuss HST FOS ultraviolet spectroscopy and
NICMOS near-infrared photometry of four young super star clusters
 in the central
 region of the irregular starburst galaxy NGC~1569.
The new observations coupled with previous HST WFPC2 photometry and
ground-based optical spectroscopy allow us to isolate and age-date
the hot and cool stellar components of these massive clusters.

We analyze the two components A1 and A2 of the brightest super star
cluster NGC~1569-A. This cluster received previous attention due to
the simultaneous presence of Wolf-Rayet stars and red supergiants.
The FOS spectra provide the first evidence for O-stars in NGC~1569-A,
indicating a young ($\leq 5$~Myr) stellar component in A1 and/or A2.
Comparison with other high-mass star-forming regions suggests
that the O- and
Wolf-Rayet stars are spatially coincident. If so, cluster A2 could
be the host of the very young O- and Wolf-Rayet population, and the
somewhat older red
supergiants could be predominantly located in A1.

The mass-to-light ratio of NGC~1569-A1 is analyzed in five 
optical and infrared photometric bands and compared to  
evolutionary synthesis models. No indications for an anomalous  
initial mass function are found, consistent with a scenario where   
this cluster is the progenitor of present-day globular clusters.  

The clusters A1 and A2 are compared to clusters B and \#30.
The latter two clusters are older and fully dominated by red
supergiants.
All four super star clusters provide 
a significant fraction (20~-~25\%) of the total
optical and near-infrared light in the central region of the galaxy.
The photometric properties of the super star cluster population in
NGC~1569 resemble those of the populous clusters in the Magellanic
Clouds.

\end{abstract}


\section{Introduction}

NGC~1569 is a well studied nearby starburst galaxy with sub-solar 
metallicity (Z$\approx 0.25 \pm 0.2$, Greggio et al. 1998 and 
referecne therein), hosting the two young super star clusters 
(SSCs) NGC~1569-A and B in its central region. 
A large body of observations at optical
wavelengths exists for
these luminous, high-density clusters, both ground-based  
(Arp \& Sandage 1985; Prada, Greve, \& McKeith 1994;
Ho \& Filippenko 1996a, b; Gonz\'alez~Delgado et al. 1997), and with HST
(O'Connell, Gallagher, \& Hunter 1994; Buckalew et al. 2000;
Hunter et al. 2000).
Spectroscopy and imaging photometry suggest a young stellar content of
blue stars  and red supergiants (RSG).

HI (Israel \& van Driel 1990), as well as CO and H$\alpha$ (Greve et al. 
1996) surveys of NGC~1569 have revealed spatially coincident
emission from neutral and ionized atomic and molecular gas, with a
pronounced hole at the location of the clusters themselves.
Such a hole is interpreted as the result of recent stellar winds and
supernova explosions which should confine the ionized gas in a shell,
such
as actually observed around NGC~1569-A (Waller 1991; Hunter, Hawley, \&
Gallagher 1993).
The supernova activity in the NGC~1569 SSCs can also provide a
significant fraction of the soft and most of the hard X-ray emission,
as measured by ASCA and ROSAT (Della Ceca et al. 1996).

Recently, the higher spatial resolution of the refurbished WFPC2
showed that cluster A is made of two components (A1 and A2,
De~Marchi et al. 1997).
Hunter et al. (2000) analyzed deep WFPC2 images and
resolved another 45  compact candidate star clusters. These clusters are
distributed across the galaxy. About one quarter of the clusters 
are located in the vicinity of NGC~1569-A. The clusters   
span a wide range of ages, between a few Myr up to 1~Gyr,
yet most of them are estimated to  be quite young, having
$B$,$V$,$I$ optical colors consistent with ages $\le$30~Myr.

While there is overall consensus that the SSCs in NGC~1569
are young, the precise age of their burst episodes is still under
debate.
This is particularly relevant for NGC~1569-A1 and A2 whose combined
spectra show features from both red supergiants and Wolf-Rayet (WR)
stars
(Gonz\'alez Delgado et al. 1997).
The simultaneous presence of the
two types of stars is not predicted by instantaneous burst models
but could be understood if the two populations originated in different
clusters.
Alternatively, star formation (SF) episodes lasting several Myr and/or 
the
formation of Wolf-Rayet stars in binary systems (e.g.,
Vanbeveren et al. 1998, Vanbeveren 1999) 
could account for their spatial coexistence. 

A first step towards resolving this issue was made by Buckalew
et al. (2000) whose HST He~II imagery demonstrated that the
WR and red supergiant light are spatially anti-correlated, 
the WR emission being concentrated on NGC~1569-A2.
In this paper we tackle this question with new 
ultraviolet (UV) and near-infrared
(IR) observations, using HST FOS spectra and NICMOS images.
In \S~2 we describe the observations and data reduction procedures.
In \S~3 we provide evidence for an O-star population in cluster A 
while in \S~4 we discuss the red stellar component. 
Then we establish an age sequence of the four most luminous super
star clusters (\S~5). The mass-to-light ratio of cluster A1 is
used to discuss the initial mass function (\S~6). The overall
properties of the cluster population are summarized in \S~7. The
conclusions are presented in \S~8.

\section{Observations and data reduction}

\subsection{Ultraviolet spectroscopy}

Ultraviolet spectra of NGC~1569-A\footnote{Throughout this paper we will
refer to the combined light of clusters A1 and A2 as NGC~1569-A.}
were obtained on 16 and 17~October 1996.
We used HST's FOS to secure H-mode spectra with the red and blue 
channels.
The FOS was pointed at R.A.(2000)~=~4:30:48.18 and 
Dec(2000)~=~+64:50:58.6,
and a 4-stage peak-up sequence with the $4.3''$, $1.0''$, $0.5''$, and
$0.3''$ apertures was performed. This sequence was done for the red and
blue channels separately. At the conclusion of each peak-up sequence,
NGC~1569-A was centered in the $0.3''$ circular science aperture with a 
precision of better than $0.1''$, as verified from an analysis of the 
individual peak-up images.
Note that the post-Costar diameter of the $0.3''$ aperture is $0.26''$.

The red-channel science observations were performed with the G190H and 
G270H
gratings, which have dispersions of 1.45 and 2.05~\AA\ per diode and 
cover 
1590~-~2312~\AA\ and 2222~-~3277~\AA, respectively.
The blue-side observations were done with G130H,
whose dispersion and spectral range are 1~\AA\ per diode and
1140~-1606~\AA, respectively.\\
Further information on the program and on the
observation strategy can be found under program 6408 at
{\tt http://presto.stsci.edu/public/propinfo.html}.

The raw data were retrieved from the HST Archive and processed with the
{\em calfos} pipeline using the recommended, best reference files. The
main
difference between the originally calibrated and the recalibrated data
is
due to the improved flat-fielding and dead-diode removal. We found
these differences to be negligibly small in our case. A strong emission
feature was found at $\sim$1960~\AA. Since no astrophysically plausible
emission line is expected around this wavelength, we interpreted this
feature as due to an intermittent diode.

Transformation from pixel to wavelength space was done with the standard
FOS wavelength scales for the three gratings. NGC~1569 has a heliocentric
velocity of $v_{\rm H}=-83$~\kms\ (Kinney et al. 1993). This is small 
in comparison with the FOS wavelength scale uncertainty of about 0.3~\AA.
Therefore we did not apply the $v_{\rm H}$ correction to the spectra 
shown below. Following the standard pipeline processing we combined the 
individual 
wavelength ranges into one spectrum covering 1200 to 3200~\AA. 
The spectrum was smoothed with a 5-pixel boxcar filter.
Since the original spectrum was oversampled by four pixels, the boxcar
smoothing almost retains the original spectral resolution. Finally, we
generated a second, rectified spectrum by division by a low-order spline
function that had been fit to line-free regions in the continuum. 

\begin{figure}
%\plotfiddle{fos_iue.ps}{15cm}{-90}{65}{65}{-260}{390}
\epsscale{0.85}
\plotone{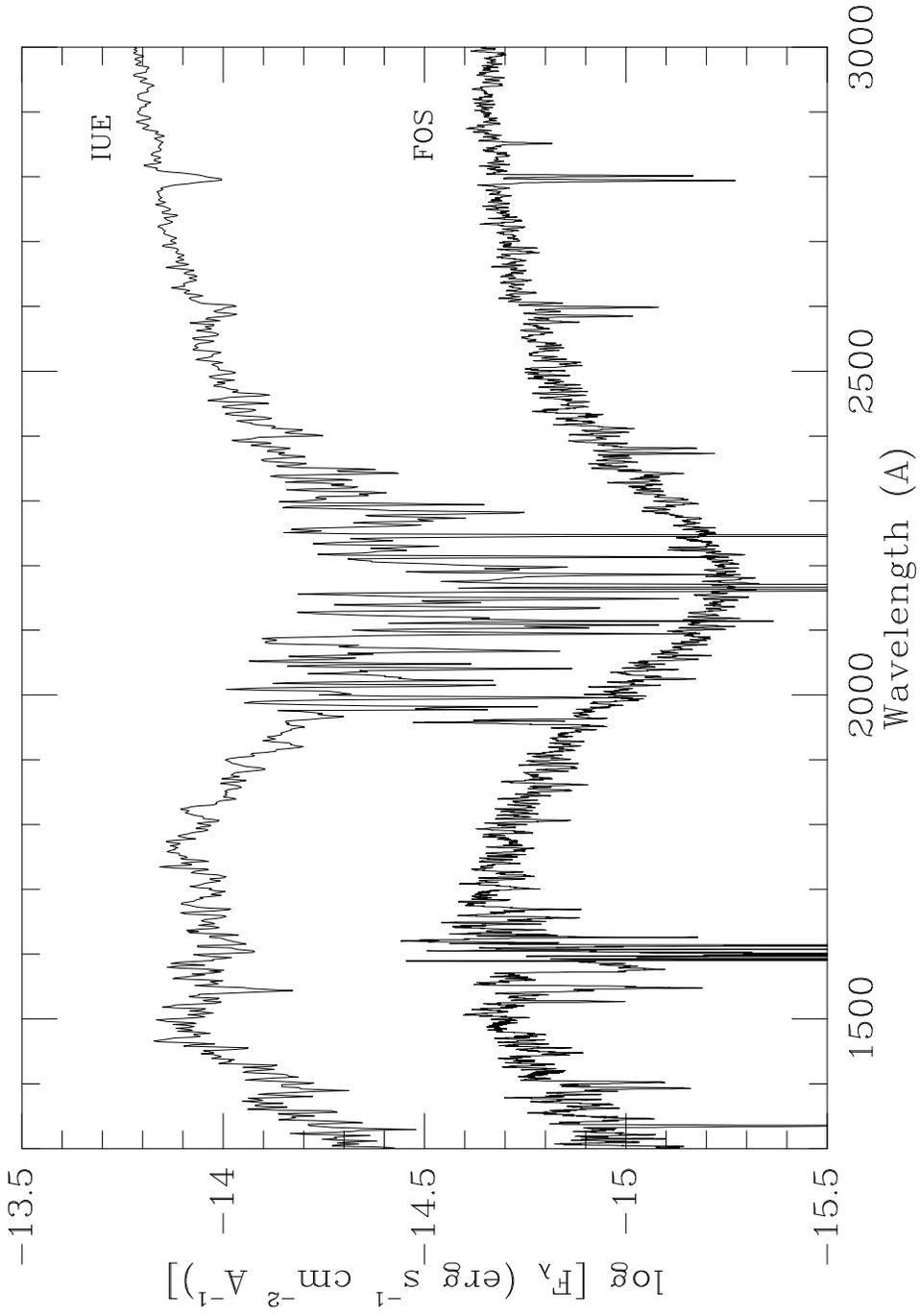}
\caption[fig]{\label{fos_iue} FOS (lower) and IUE (upper) spectra of
NGC~1569.
The FOS spectrum was taken with a circular aperture of diameter $0.26''$,
and includes the total light from the super star cluster NGC~1569-A.
The IUE aperture
has dimensions $10'' \times 20''$ and covers most of the central region.
The continuum shapes of the two spectra are quite similar. The
lower spectral resolution of IUE accounts for most differences in the
line
spectrum.}
\end{figure}

The flux-calibrated FOS spectrum of NGC~1569-A is plotted in
Fig.~\ref{fos_iue}.
The FOS aperture has a diameter of
$0.26''$, sampling most of the cluster luminosity,
taking also into account the photometric
correction included in the pipeline to compensate for 
Point Spread Function (PSF) light losses.

The broad absorption
at 2200~\AA\ is due to strong foreground extinction. The high
noise level around 1600~\AA\ is in the overlap region between 
the G130H and G190H spectra where the detector sensitivity is low.
In Fig.~\ref{fos_iue} we also plot for comparison 
an IUE spectrum of NGC~1569,  
discussed by McQuade, Calzetti, \& Kinney (1995).
The IUE aperture has dimensions $10'' \times 20''$, covering
the whole central SF region in this galaxy.

At wavelengths below
about 2500~\AA\ the shapes of the FOS and IUE spectra are essentially
identical, suggesting a spatially uniform foreground and intrinsic
reddening within the galaxy and negligible aperture effects.
The flux level of the FOS data is lower
by about a factor of 5 in comparison with IUE.
The IUE spectrum becomes redder than  the FOS spectrum at the longest
wavelengths plotted in Fig.~\ref{fos_iue}. We interpret this flux
excess as due to a surrounding, somewhat older population, which is
included in the IUE aperture.

The de-reddened FOS spectrum of NGC~1569-A is shown in Fig.~\ref{lum}.
Conversion to luminosity units was done for an assumed distance of 
$d=2.2$~Mpc (De Marchi et al. 1997).
We used the strength of the 2200~\AA\ feature to obtain an independent
estimate of the Galactic contribution to $E(B-V)$. Adopting the reddening
curve of Mathis (1990), we varied $E(B-V)$ to remove the 2200~\AA\ 
depression.
$E(B-V)=0.55 \pm 0.05$ was found, in good agreement with the value
quoted in the literature.
The additional intrinsic component of $E(B-V)$ was derived from a 
comparison 
of the observed spectral slope (after correction for foreground 
reddening) 
and the theoretical prediction $F_\lambda \propto \lambda^{-2.5}$ 
(Leitherer \& Heckman 1995). We adopted the starburst obscuration law of 
Calzetti (1997) to derive an intrinsic reddening of $E(B-V)=0.15 \pm 
0.05$. 
The total reddening then becomes $E(B-V)=0.7 \pm 0.1$.

\subsection{Near-infrared photometry}

HST NICMOS IR imaging of NGC~1569 with the F110W and F160W filters
using the NIC2 camera centered in between the SSCs A and B 
was obtained on 25 February, 1998.
We performed a spiral dithering  with a step size of $0.2''$
for an on-source total integration time of about 85~min in each filter.
The dithered frames have been combined using a drizzling technique,
and the final images with $0.0375''$ pixel$^{-1}$ have been photometrically
calibrated by transforming the count rates into the VEGAMAG system,
according to the PHOTNU and zero point values reported
in the NICMOS web page. 

Details of the photometry of the stellar population, with an extensive 
discussion of the reduction procedure used to correct the frames for the 
instrumental response are presented by Aloisi et al. (2001). 
Here we focus on the cluster photometry. 
The main objective of our study is the double cluster
NGC~1569-A but we include the clusters B and \#30 as well. The latter
cluster was identified by Hunter et al. (2000) and has properties 
similar
to those of A and B.
By combining the NICMOS and WFPC2 data, we perform a multicolor analysis 
and are able to characterize the different  blue and red stellar 
components 
of each cluster.

The SSCs are barely resolved due to the width of
the NIC2 PSF.
We used ROMAFOT (Buonanno et al. 1983) to fit each cluster luminosity
profile with a Moffat function after subtraction of the local 
background. 
With such a technique we can also
properly deconvolve the two components in the double cluster NGC~1569-A.

We applied this procedure to both the NIC2 F110W, F160W  and the
PC F380W, F439W, F555W images (De~Marchi et al. 1997; Greggio et al.
1998).
We also performed aperture photometry within a radius
of $\le$0.9" and the resulting magnitudes turn out to be $\le$0.2 mag
brighter than those obtained with the profile fitting technique.
Finally, we compared our optical magnitudes obtained with the
fitting technique with those by De~Marchi et al. 
(1997, clusters A1, A2, and B) and Hunter et al.
(2000, clusters B and \#30) obtained from aperture photometry and 
once again we found that the latter values are a few tenths of a
magnitude brighter.
Although our profile fitting technique might somewhat underestimate the 
total cluster luminosity, it should however provide more reliable colors for 
an analysis of the cluster stellar populations since it is
less contaminated by the galaxy field and/or foreground stars. 
We assume 0.2~mag
as a conservative error in the derived magnitudes and colors.
This value takes into account the  
uncertainty of the aperture effects and background subtraction.

Due to its low Galactic latitude of $b=11.2^\circ$, Galactic
foreground extinction is significant in NGC~1569.
In Table~\ref{colors} the dereddened F555W$_0$ magnitude and
(F380W-F439W)$_0$,
(F439W-F555W)$_0$,
(F110W-F160W)$_0$,
(F555W-F110W)$_0$, 
(F439W-F160W)$_0$
colors of the four SSCs are listed, assuming $A_{\rm F555W}=1.73$,
$E(F380W-F439W)=0.30$,
$E(F439W-F555W)=0.56$, 
$E(F110W-F160W)=0.28$,
$E(F555W-F110W)=1.18$, and
$E(F439W-F160W)=2.02$,
according to the values quoted by Holtzmann et al. (1995, their
Table 12a,b) and Calzetti (private communication).  \\
If a further intrinsic reddening of about 0.15~mag in $E(B-V)$
is taken into account (see \S~2.1), the
(F555W-F110W)$_0$ and (F439W-F160W)$_0$ colors
become $\sim$0.2 and 0.3~mag bluer, respectively,
whereas the (F380W-F439W)$_0$ and the 
(F110W-F160W)$_0$ colors are not significantly affected.
In Fig.~\ref{energy} we plot the energy distribution of each SSC,
according to the HST photometry of Table~\ref{colors} and the
zero-magnitude flux conversion reported by Holtzmann et al. 
(1995, cf. their Table 9) and on the NICMOS web page.

We also transformed our HST colors into the ground-based Johnson
$U$,$B$,$V$,$J$,$H$
system with the relations described by Origlia \& Leitherer (2000).
Using these transformations we found
that (F380W-F439W)$_0=-0.5$ corresponds to $(U-B)_0\approx -1.1$,
while (F380W-F439W)$_0=-0.1$, measured in cluster \#30, corresponds
to $(U-B)_0 \approx -0.3$.
(F439W-F555W) is very similar to  $(B-V)$ for (F439W-F555W)$<0.8$, 
(F110W-F160W)$_0=0.8 -0.9$ corresponds to $(J-H)_0 \approx 0.5-0.6$, 
while
(F110W-F160W)$_0=0.3-0.5$ implies 0.1~-~0.2~mag bluer $(J-H)_0$ 
colors.

\section{Evidence for an O-star population}

\begin{figure}
%\plotfiddle{n1569_lum.ps}{18cm}{0}{75}{75}{-230}{-20}
\epsscale{0.45}
\plotone{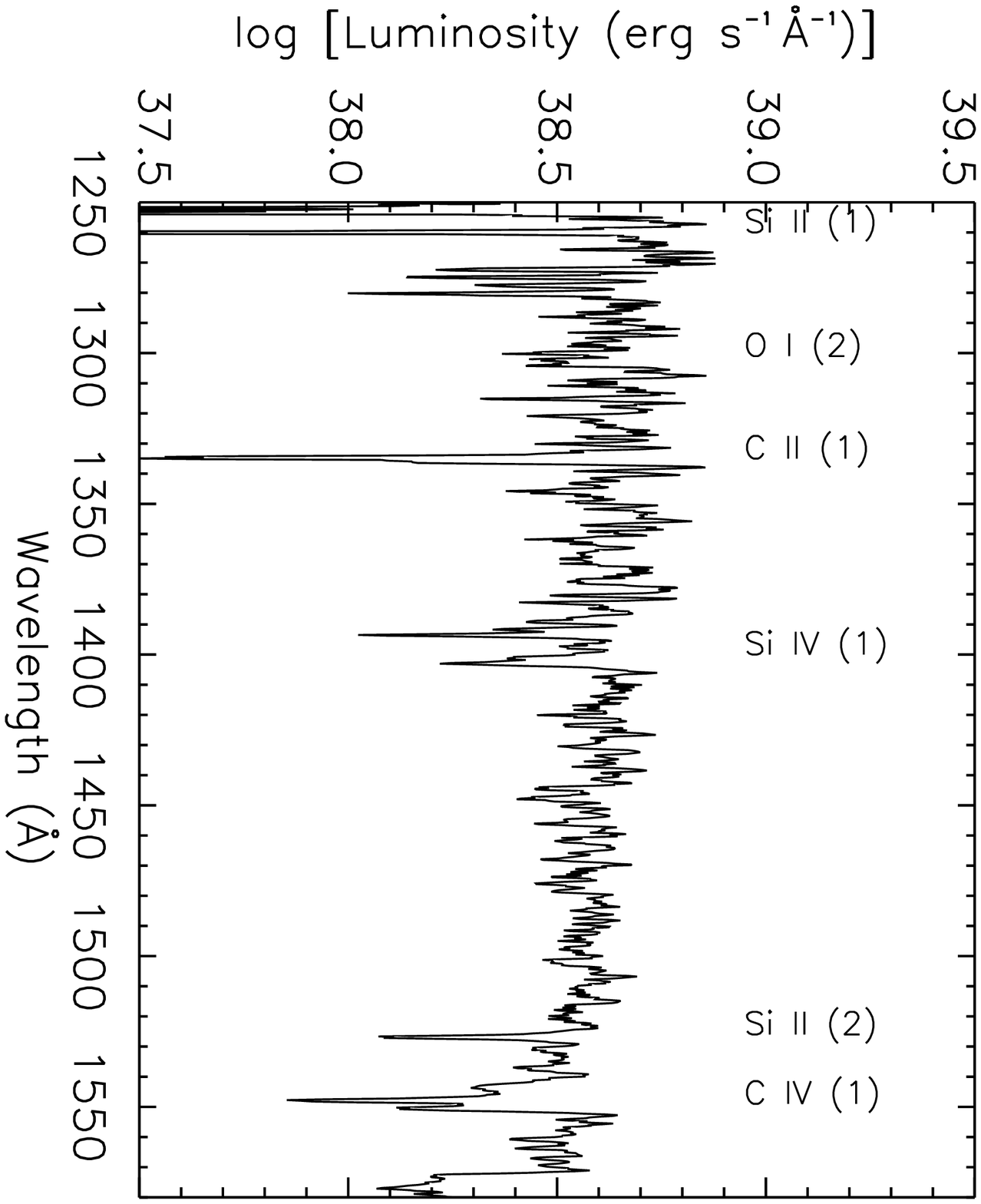}
\plotone{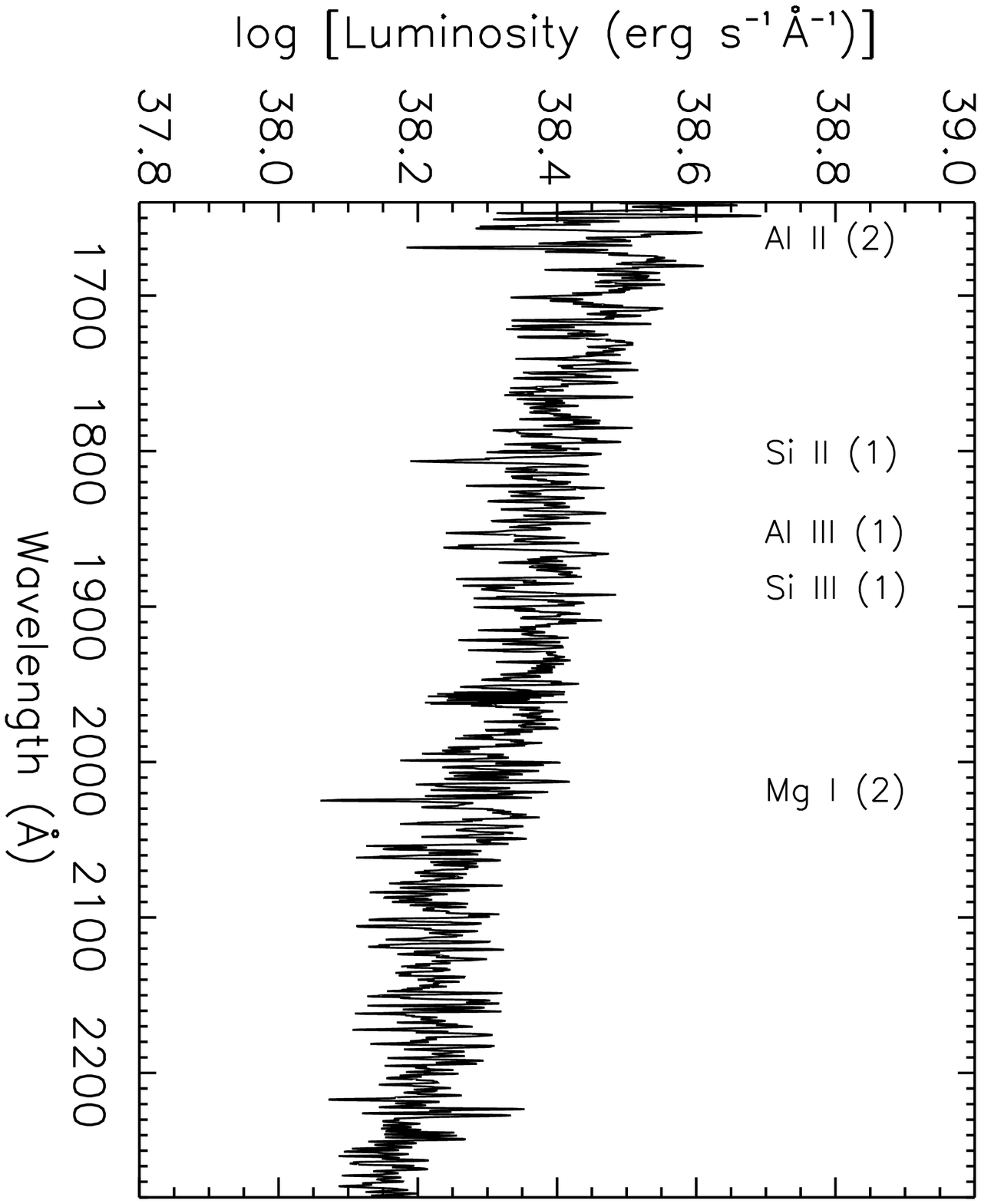}
\plotone{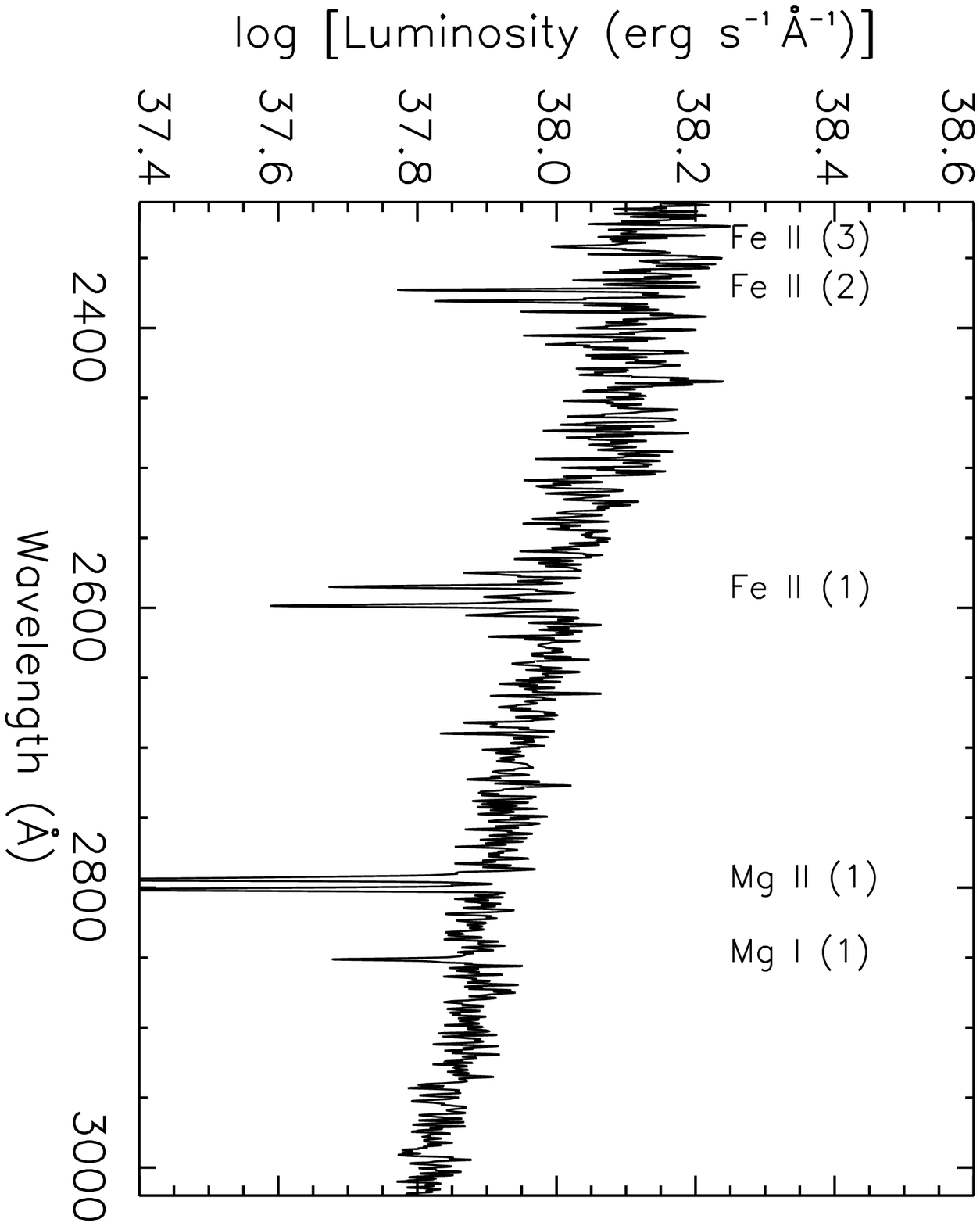}
\caption[fig]{\label{lum} FOS spectrum of the super star cluster
NGC~1569-A in luminosity units ($d=2.2$~Mpc; $E(B-V)=0.70$).
The strongest spectral lines are identified using Moore's (1950) nomenclature. The
region around 1960~\AA\ was manually edited to remove an emission feature due
to an intermittent diode.
}
\end{figure}

\begin{figure}
%\plotfiddle{energy.ps}{15cm}{0}{75}{75}{-230}{-120}
\epsscale{1.00}
\plotone{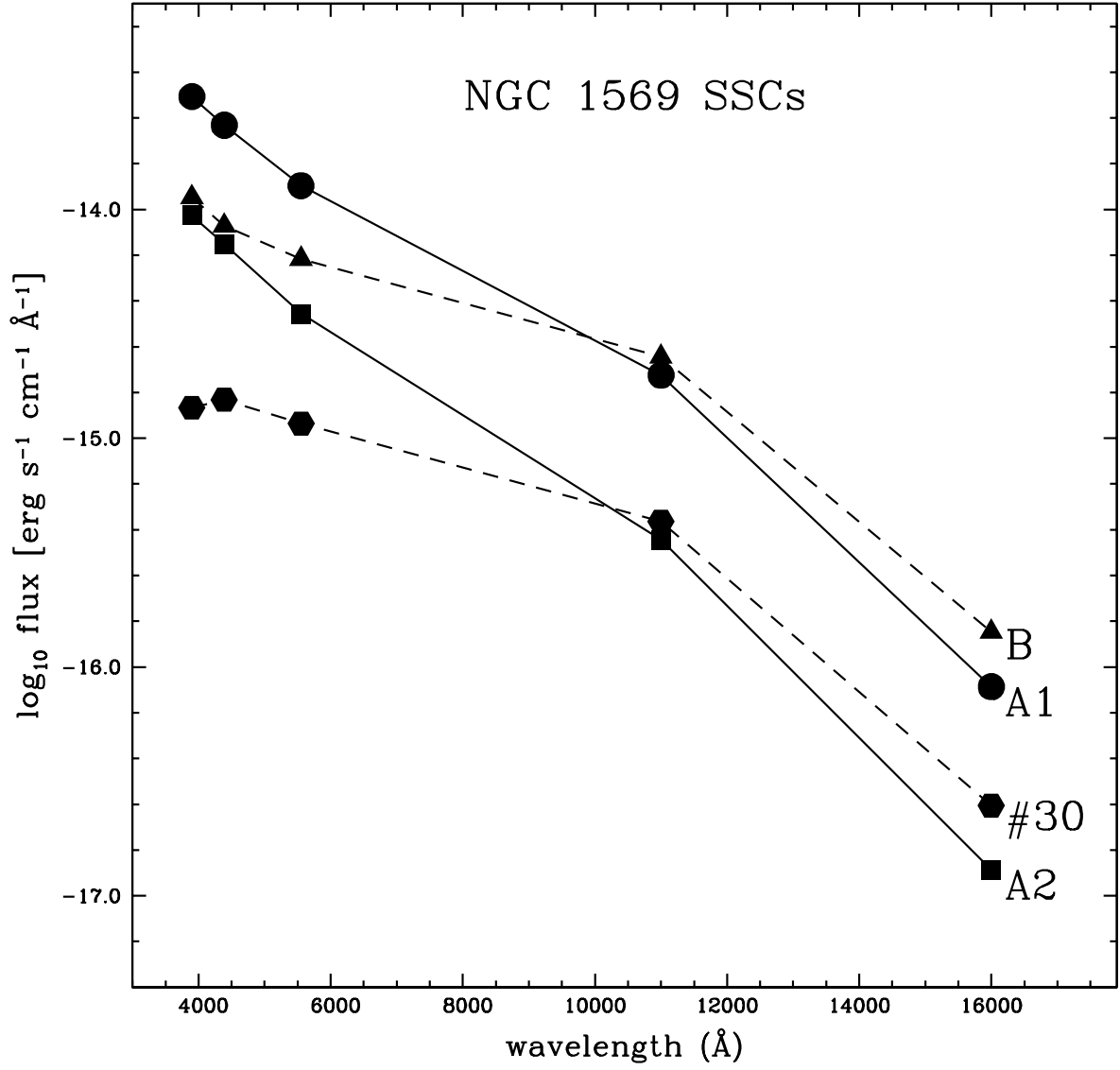}
\caption[fig]{\label{energy}
Dereddened energy distributions for the four super star clusters in
NGC~1569, as derived from our HST optical and near-IR  photometry.}
\end{figure}

The strongest lines in the UV spectrum in Fig.~\ref{lum}
 are all ground-level
transitions, suggesting that they
originate in stellar winds, interstellar gas, or in the
Milky Way halo, but have little photospheric contribution. The spectral
resolution of FOS is insufficient to resolve any Milky Way halo
absorption
(if present) from absorption in NGC~1569 due to the small heliocentric
velocity of NGC~1569 ($v_{\rm H} =-83$~\kms).
This must be kept in mind during the following discussion.

The wavelength zero point of the G130H spectrum can be determined from
the strong geocoronal Ly$\alpha$. The observed velocity is +143~\kms. 
This
offset corresponds to a shift by half a diode, which is within the 
expectation.
Assuming that geocoronal Ly$\alpha$ should be unshifted, we corrected
the G130H wavelength scale by -143~\kms\ and by $-v_{\rm H}$ and 
determined
the velocities of \cii, \siiv, \siii, and \civ. 
In the case of \siiv\ and \civ\ we ignored the blue wings due to stellar
winds which are seen shortward of the narrow interstellar absorptions.
The mean velocity is
$(-46 \pm 50)$~\kms, consistent with no velocity shift. Since the G190H
and G270H observations were done with an independent target acquisition
several hours after the G130H observations, the velocity offset 
determined in
G130H does not apply to G190H and G270H. Heliocentric velocities 
(without
applying any other correction) were derived for
Al~II $\lambda$1670,
Si~II $\lambda$1808,
Al~III $\lambda$1858,
Mg~I $\lambda$2024,
Fe~II $\lambda$2378,
Fe~II $\lambda$2593,
Mg~II $\lambda$2800, and
Mg~I $\lambda$2852. We found a mean velocity of $(+64 \pm 52)$~\kms.
This agrees with the G130H result; a zero point error of order 100~\kms\ 
is
expected for these red-channel observations as well. Note that the zero
point error is {\em not} caused by centering errors of the target but by
the non-repeatability of the grating wheel. 

The relatively low spectral resolution of FOS, combined with possible 
blending
by Milky Way halo lines precludes any quantitative interpretation of the
measured velocities. Heckman \& Leitherer (1997), using HST's GHRS,
detected a blueshift of all interstellar lines in the
starbursting dwarf galaxy NGC~1705 by 80~\kms, and
ascribed it to an outflowing galactic superwind. 
Such a blueshift could well be present in the spectrum
of NGC~1569-A but it would remain undetected in our data.

The quality of the data is, however, adequate to address a central 
question
we would like to answer: Is there evidence for a young O-star population
in the SSC NGC~1569-A?
O stars have powerful stellar winds whose associated
broad absorptions and P~Cygni profiles are detectable in young 
populations
(Robert, Leitherer, \& Heckman 1993; Leitherer, Robert, \& Heckman 1995;
de~Mello, Leitherer, \& Heckman 2000). Inspection of Fig.~\ref{lum}
suggests that hot-star winds are indeed seen in NGC~1569-A. 
\siiv\ and \civ\ have broad, blueshifted absorptions superimposed on the 
narrow interstellar lines.
They are clearly broader than the purely interstellar lines of, e.g.,
\siii\ or \cii.
Note that \siiv\ and \civ\ have  blueshifted wings
 but we were not able to decompose the
stellar and interstellar components since they are unresolved.

\begin{figure}
%\plotfiddle{n1569_siiv.ps}{15cm}{-90}{65}{65}{-260}{390}
\epsscale{0.85}
\plotone{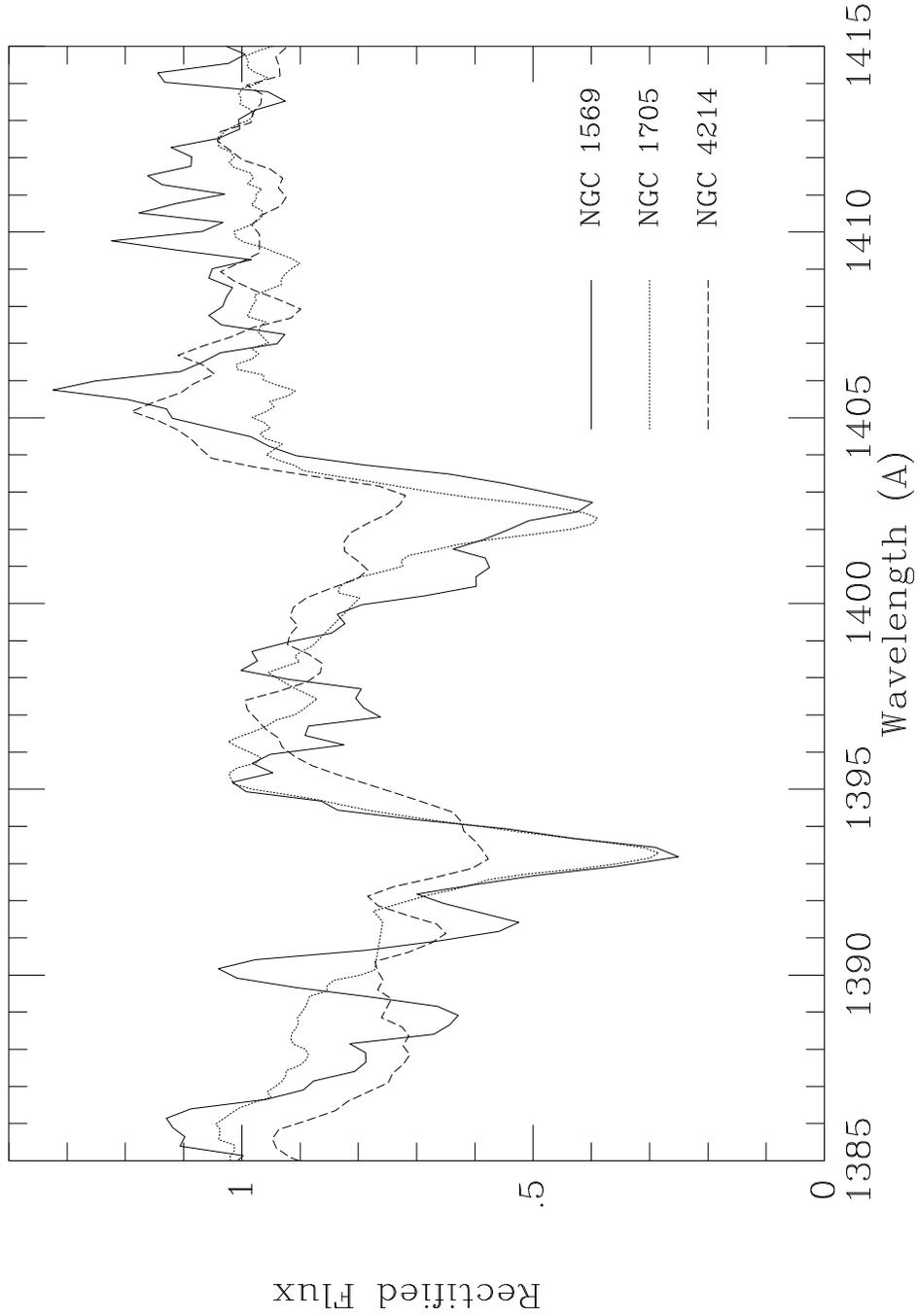}
\caption[fig]{\label{siiv} \siiv\ in NGC~1569-A (solid line) in comparison
with observations of NGC~1705-1 (dotted) and NGC~4214 (dashed). The
profile
in NGC~1569-A shows a stellar-wind feature whose strength is
intermediate
between that seen in the other two galaxies. (A strong Milky Way halo
absorption doublet
in NGC~1705-1 at 1391 and 1399~\AA\ was removed for clarity reasons.)
}
\end{figure}

\begin{figure}
%\plotfiddle{n1569_civ.ps}{15cm}{-90}{65}{65}{-260}{390}
\epsscale{0.85}
\plotone{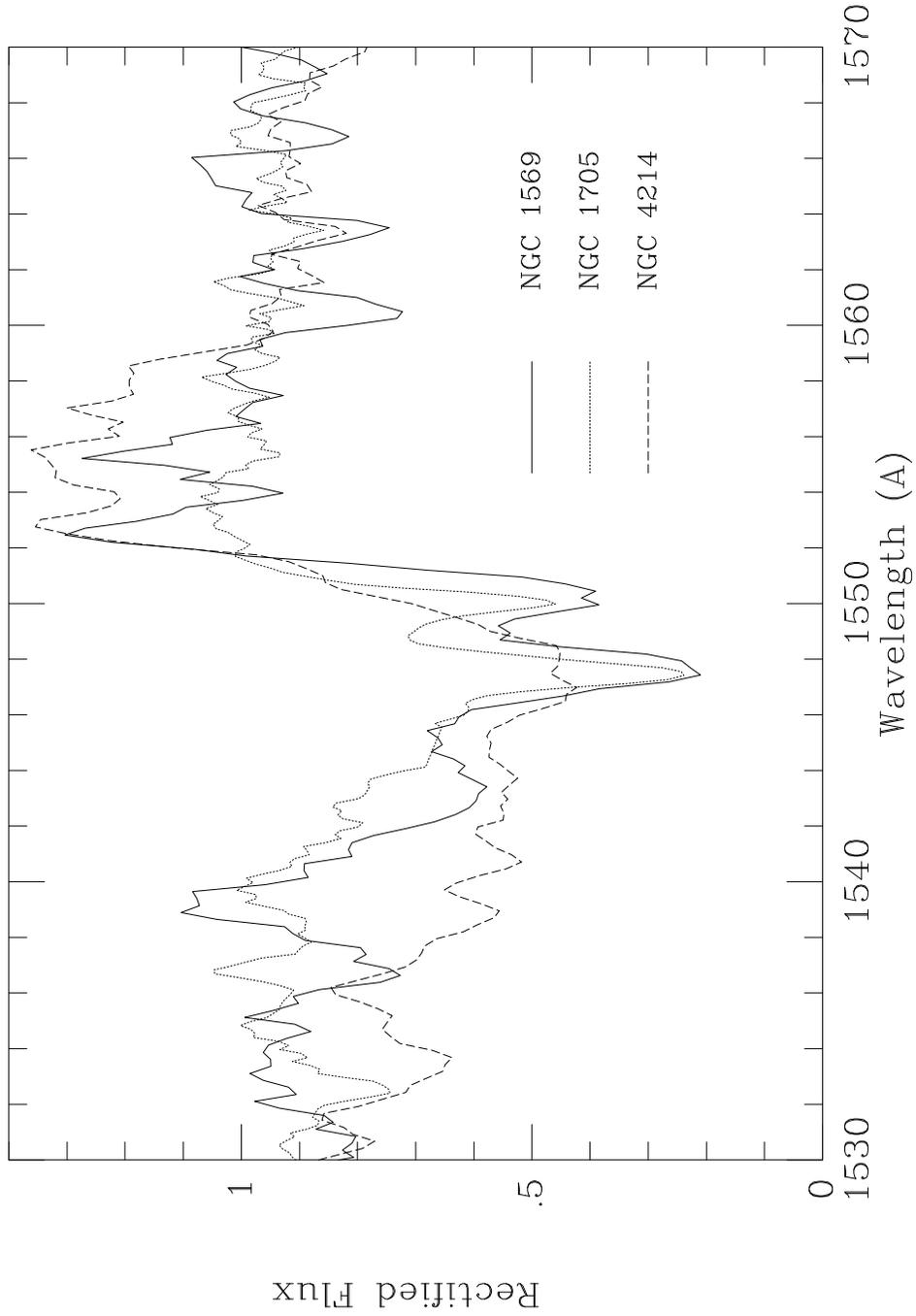}
\caption[fig]{\label{civ} Same as Fig.~\ref{siiv} but for \civ. As 
we did for \siiv, the strong Milky Way halo line at 1545~\AA\
in NGC~1705-1 was removed.}
\end{figure}

The case for an O-star component becomes more compelling when
the NGC~1569-A data are compared to UV spectra of stellar populations
with and without O component. This is done in Figs.~\ref{siiv} and
\ref{civ}, which show the spectral regions around \siiv\ and 
\civ, respectively. 
The NGC~1705-1 data were taken with the GHRS (from Heckman \& Leitherer 
1997) and refer to a 10 to 15~Myr old cluster with no O stars. 
In this case, \siiv\ and \civ\ are narrow due to interstellar and weak
photospheric contributions. NGC~4214 is a very young starburst cluster
discussed by Leitherer et al. (1996). Its FOS spectrum shows  strong
P~Cygni profiles due to O-star winds. NGC~1569-A is intermediate between
the two cases. While its P~Cygni profile is not as strong as in 
NGC~4214,
excess wind emission and absorption with respect to NGC~1705-1 is 
clearly detected.  NGC~1569
is a most unfavorable case for the application of ultraviolet 
synthesis models. The high foreground reddening and low systemic
velocity lead to dominating Milky Way absorption almost coinciding with
the stellar wind lines at the resolution of the FOS. Therefore tailored
synthesis models would not be appropriate. Nevertheless we can attempt
to narrow down the age range of the observed O-star population with
the models of Leitherer et al. (2001). Using their Figure~6, we find
that the blueshifted absorptions and the emission components in
\siiv\ and \civ\ suggest ages between 4 and 7~Myr
if the SF occurred instantaneously.

NGC~1569-A and NGC~1705-1 are two SSCs which were previously
thought to have similar age, based on their optical colors (O'Connell,
Gallagher, \& Hunter 1994). The UV spectra indicate significant
differences:
NGC~1705-1 has no O-star component, consistent with an age above
10~Myr.
NGC~1569-A, on the other hand, does have a weak O population.

\section{The red stellar population}

The UV data are sensitive to the hottest, youngest population. 
The presence or absence of an additional cooler and older population can 
be probed with the longer-wavelength data in Table~\ref{colors} and  
Fig.~\ref{energy}.

A1, A2, and B have identical (F380W-F439W)$_0$ colors, while
cluster \#30 is 0.4~mag redder. Both clusters B and \#30 have
slightly redder (F439W-F555W)$_0$ than A1 and A2.
The IR (F110W-F160W)$_0$ and optical-to-IR      
(F555W-F110W)$_0$ and (F439W-F160W)$_0$
colors show more variations.
NGC~1569-A2 is the bluest cluster,
NGC~1569-B is the reddest and the brightest in the IR, while
NGC~1569-A1 is the brightest in the visual range. The colors suggest
that A1 and A2 are bluer, and therefore younger than B and \#30.

Theoretical colors of evolving starbursts are quite sensitive to cool
stars. Starburst99 models (Leitherer et al. 1999) predict a sudden
reddening after about 6 to 8~Myr, depending on metallicity. As
discussed by Origlia et al. (1999), the predictions are uncertain
due to deficiency in the stellar evolution models of RSGs at low
metallicity. Nevertheless, a safe prediction at any metallicity is that
a redder color indicates a cooler stellar population and that the
cool population has a more significant effect towards longer
wavelengths.
Using the Starburst99 colors for guidance, we expect $(U-B)$ and $(B-V)$
to vary less than 0.5~mag between $t =1$~Myr and 30~Myr, i.e. between
the O-star and RSG dominated phases. This barely exceeds the photometric
uncertainties and is therefore less useful in our particular case. The
optical-to-IR $(V-J)$ and $(V-H)$ colors, however, are expected to
change by up to 2~mag.

The photometry in Table~\ref{colors} suggests that (i)
clusters B and \#30 are in a RSG dominated phase, (ii) A2 has few (if
any) RSGs, and (iii) A1 is intermediate between A2 and B/\#30.

Ground-based red and near-IR spectroscopy of clusters A and B show
the Ca~II $\lambda$8600 triplet as well as the CO $\lambda$2.3 feature
in absorption (Prada et al. 1994; Gonz\'alez~Delgado et al. 1997;
Hunter et al. 2000), indicating the presence of red supergiants.
Ho \& Filippenko (1996a, b) secured  high-resolution spectra of cluster
A between $\sim$5000 and 6300~\AA. Numerous weak absorption lines
from RSGs allowed them to estimate a stellar velocity dispersion of
15.7~\kms.
Moreover, many Balmer and Paschen absorption lines
(Arp \& Sandage 1985; Prada et al. 1994; Gonz\'alez~Delgado et al. 1997)
have been detected in this cluster.
These lines are common features in A-type stars and,
in the case of the NGC~1569 SSCs, are most probably associated with
luminous supergiants (Arp \& Sandage 1985) rather than with an older
($\approx$1 Gyr) main-sequence population.

\section{The relative ages of the clusters}

Prada et al. (1994)
classified the SSCs A and B in the age sequence proposed by 
Bica, Alloin \& Santos  (1990) for the LMC clusters, which is based
on the CaII triplet and Paschen line equivalent widths.
They suggested that NGC~1569-A is slightly older than NGC~1569-B,
and assigned NGC~1569-A to their Y$_c$ (12-35~Myr) and NGC~1569-B to 
their Y$_b$ (7-12~Myr) groups. However, even LMC clusters in the 
Y$_a$ (5-7~Myr) group have CaII triplet equivalent widths very close to 
that observed in NGC~1569-A (Hunter et al. 2000).
Therefore NGC~1569-A can be very young as well, with luminous O-B stars 
starting to evolve into massive RSGs.

Gonz\'alez~Delgado et al. (1997) measured Balmer jumps $\le$1.15 
in both cluster A and B.
Instantaneous burst models at metallicity below solar 
(but to some extent also continuous SF models) 
predict a very young ($<5$~Myr) age if $(U-B)\approx -1$ and
Balmer jumps$\le$1.15 are observed
(Fig.~13 of Leitherer \& Heckman 1995 and
Fig.~5 of Gonz\'alez~Delgado et al. 1997).
This appears to be too young to allow the simultaneous presence of hot
stars and RSGs.

Nevertheless, Origlia et al. (1999) show that
the $(U-B)$ colors predicted by Leitherer \& Heckman (1995)
using the Geneva evolutionary tracks at subsolar metallicity
are too red compared to those observed in star clusters in the
Magellanic Clouds. 
As a possible explanation they suggest that these tracks 
overestimate the lifetime
in the blue portion of the loops during the core He-burning phase.  
In such blue excursions the massive stars are cooler (A-type stars) 
than the O-B main sequence population and tend to redden the 
blue portion of the optical spectrum.
Shorter blue loop lifetimes can 
indeed produce bluer $(U-B)$ colors (a few tenths of a magnitude) and 
shallower ($\sim$20\%) Balmer jumps for a given age since the 
contribution to the blue luminosity by A-type stars is reduced. 

If this bias is taken into account in the models,
the measured $(U-B)$ color and
Balmer jump in NGC~1569-A and B can be also consistent with somewhat 
older ages at which RSGs are actually predicted to be observed.
$(U-B)_0\approx -1$ is indeed typical of the youngest LMC and SMC clusters
(e.g., van den Bergh 1981),
which are known to contain red supergiants, on the basis of the
strong CaII triplet, CO lines and the red $(J-H)_0$ color
(Bica, Alloin \& Santos 1990; Persson et al. 1983; Oliva \& Origlia 1998).

Gonz\'alez~Delgado et al. (1997) also detected the
$\lambda$4686 WR bump in cluster A and no $\lambda$5808 WR bump, 
indicating that these WR stars should be of the WN subtype.
The equivalent width of the WR feature is 3.4~\AA, while 
the inferred WR/WR+O ratio is $\ge$10\%.

Continuous SF models predict a WR bump equivalent width $\le$1~\AA\ and
a corresponding WR/WR+O value $\le$~2\%
at metallicities below solar
(Fig.~6 of Gonz\'alez~Delgado et al. 1997; Figs. 2 and 4 of Cervi\~no 
\& Mas-Hesse 1994). 
These values are well below the corresponding quantities measured in 
cluster A, so continuous SF does 
not seem adequate to explain the coexistence of both WR stars and RSGs 
in this cluster. 

The coexistence of WRs and RSGs can be predicted by the 
so called binary channel scenario (Vanbeveren et al. 1998; Vanbeveren 
1999) 
for the formation of the WR population. 
Stars with initial masses $\ge$40 \Ms\ undergo mass loss mainly due to 
the strong stellar winds (the single-star channel; e.g.; Maeder \& 
Meynet 1994 and references therein),
while Roche lobe overflow in massive close binaries occurs at 
lower initial masses.
This implies that the binary channel continues to form WR stars at 
later epochs than the single channel, and it extends the lifetime of
the global WR population, since its progenitors can have
initial masses smaller than the WR mass limit for single stars.

Recent models by Schaerer \& Vacca (1998) 
using evolutionary tracks with enhanced mass loss (Meynet et al. 1994) 
and a WR binary frequency of 20\%, show that there is a major age 
separation at $\sim$5 Myr between the single and binary channels. 
As a first order approximation the WR bump equivalent width 
scales proportionally to the binary frequency if the latter is not 
too large.  
According to the models of Schaerer \& Vacca (1998), 
equivalent widths of 3-4~\AA\ in the  $\lambda$4686 WR bump 
can be easily obtained at ages $\le$10 Myr for binary frequencies $\le$50\%. 
Gonz\'alez~Delgado et al. (1997) used the Geneva evolutionary 
tracks with standard mass loss and obtained 
an equivalent width of 3.4~\AA\ slightly later, around 14~Myr,
for the same binary frequency.
Nevertheless, it is rather well established  that the lower the mass 
loss, the later the appearance of the WR phase (Maeder \& Meynet 1994).
Therefore the binary channel seems to be able to 
account for spatially coincident and coeval ($\le$10 Myr) 
WR and RSG populations, even though with a relative high 
binary frequency and some fine tuning for the burst age.
 
On the other hand, as discussed in \S~4, there is strong observational 
evidence that cluster A is double and A1 is redder than A2. 
This constraint favors a scenario 
where cluster A1, slightly redder, contains RSGs, 
while A2, the bluest, contains the WR stars and the O-star population 
(cf. \S~3), which can be likely coeval.
If the blue and red populations are spatially separated, the single 
channel scenario for the formation of the WR stars is favored since 
in order to have only WRs and not RSGs the burst age should be
$\le$5 Myr. At this stage the binary channel is not yet fully active
(Schaerer \& Vacca 1998).

To summarize, the existing body of observations is consistent with the
age sequence derived from the near-IR photometry:
B and \#30 are the most evolved clusters ($\ge$10 Myr), A2 is the
youngest ($\le$5 Myr), and A1 is intermediate.

\section{Constraints on the initial mass function}

The initial mass function (IMF)
is of considerable interest for the interpretation of SSCs as
progenitors of present-day globular clusters. If low-mass stars are
not present in significant numbers, SSCs are unlikely to evolve into
old globular clusters, either because their dynamical survival times
would be too short or because they would simply fade on a cosmologically
short time scale.
The observed mass-to-light ratio ($M/L$) allows constraints on the IMF 
since the luminosity is dominated by the high-mass end, and the mass
by the low-mass end of the distribution.
Due to their low luminosity, low-mass stars can only be inferred via 
their gravitational forces, e.g., by measuring the velocity dispersion 
of the cluster. This experiment was done by Ho \& Filippenko (1996a, b) 
who resolved the RSG features in the optical spectrum of NGC~1569-A.
Their ground-based spectra could not separate the two components of this 
double cluster but since A1 is brighter than A2 one can reasonably
assume that the spectrum is dominated by the former.

Sternberg (1998) analyzed the visual mass-to-light ratio,
$M/L_{\rm V}$, using the mass derived from Ho's \& Filippenko's
measurement.
Since the crossing time scale is much shorter than its age, the cluster
is expected to be dynamically mixed,  and the virial theorem can be used
to infer the dynamical mass. Using the same methodology as Sternberg,
but adopting a distance of 2.2~Mpc and a half-light radius of
1.6~pc (De Marchi et al. 1997), 
we infer a dynamical mass of $2.7 \times 10^5$~\Ms. Our new
near-IR data allow us to extend the mass-to-light ratio analysis to the
wavelength region where RSGs are brightest. We transformed
the luminosities  from our HST optical and IR photometry
(Table~1)  into the standard Johnson system and adopted a conservative
error of a factor of 2 in either direction for the $M/L$ ratio in each
passband. The error is dominated by the systematic uncertainties of the
cluster mass. In order to be able to perform a comparison with
evolutionary synthesis models, we need to assume an age for cluster A1.
We assign conservative upper and lower
limits of 15 and 5~Myr, respectively. The upper limit comes from the
presence of the CO feature (Hunter et al. 2000), which could bring the
age of A1 close to that of B, and the lower limit is imposed by the
occurrence of the first RSGs according to stellar-evolution models.

\begin{figure}
%\plotfiddle{movl.ps}{15cm}{0}{80}{80}{-230}{-120}
\plotone{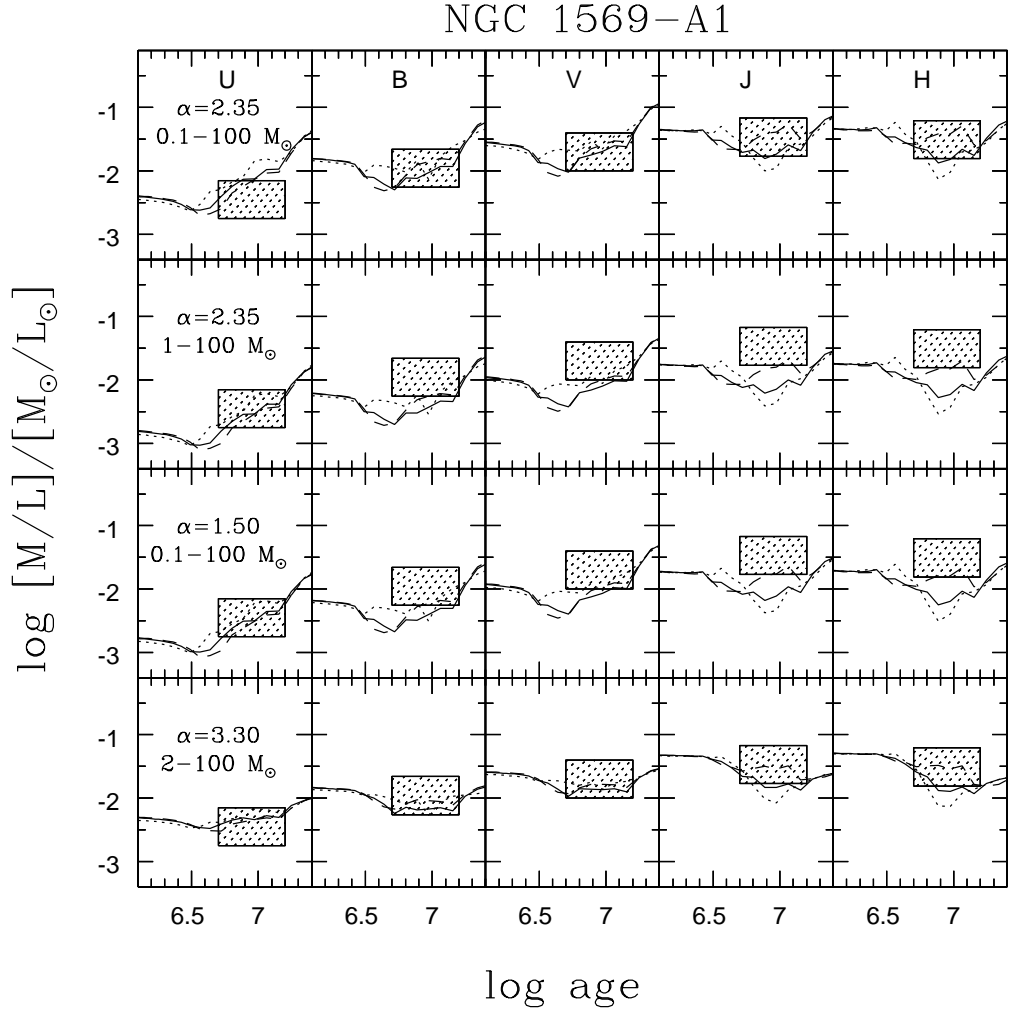}
\caption[fig]{\label{movl}
Mass-to-light ratios vs. time in the $U$,$B$,$V$,$J$,$H$ photometric
bands
predicted by evolutionary synthesis models with
different IMF parameters, as indicated in each left panel. The models
are
for a metallicity of \Zs\ (dotted), 1/4~\Zs\ (solid), and  1/10~\Zs
(dashed).
Rectangles indicate the measured value in NGC~1569-A1.}
\end{figure}

In Fig.~\ref{movl} we compare the derived mass-to-light ratios 
to those expected from evolutionary synthesis models by
Leitherer \& Heckman (1995). The panels in this figure show the
models for four different IMFs and for metallicities of \Zs,
1/4~\Zs, and 1/10~\Zs. The 1/4~\Zs\ models should be most appropriate
for NGC~1569 but we provide models with higher and lower metallicity
as a guide for the expected uncertainties,
since evolutionary tracks do not properly account
for metallicity effects in RSGs (cf. Origlia et al. 1999).

We explored a variety of IMF parameters with the goal of constraining 
the
relative fraction of low-mass stars. The IMF is approximated by a power
law of the form $N(M) \propto M^{-\alpha}$ between the adopted lower and
upper mass limits, \Mlow\ and \Mup, respectively. The low-mass star
contribution is increased by increasing $\alpha$ and/or by
lowering \Mlow. A Salpeter IMF has $\alpha=2.35$ in this
nomenclature. Fig.~\ref{movl} illustrates four examples:
models with a standard Salpeter's IMF (top panel),
models with a deficit in low-mass stars resulting either from a
truncated or a flatter IMF compared to the standard Salpeter's 
(central panels) and models with a low-mass truncation and a steeper IMF
slope (bottom panel). The best fits to the data have been obtained
with a standard Salpeter IMF with lower mass cutoff
down to the hydrogen-burning limit or an IMF truncated at
$\approx2~M_{\odot}$ and with a very steep slope over the whole range of
masses, but this latter IMF is not commonly observed in stellar
populations. Hence, even accounting for the uncertainties in the
evolutionary models, Fig~\ref{movl} generally does not support an 
IMF deficient in low-mass stars. 
The strongest constraint comes from the $J$ and $H$ luminosities, 
which sample RSGs.
These near-IR data support the previous result of Sternberg (1998) that
NGC~1569-A has a normal IMF with no deficit in low-mass stars.

\section{Large-scale properties of the cluster population}

\begin{figure}
%\plotfiddle{integr.ps}{15cm}{0}{80}{80}{-230}{-120}
\plotone{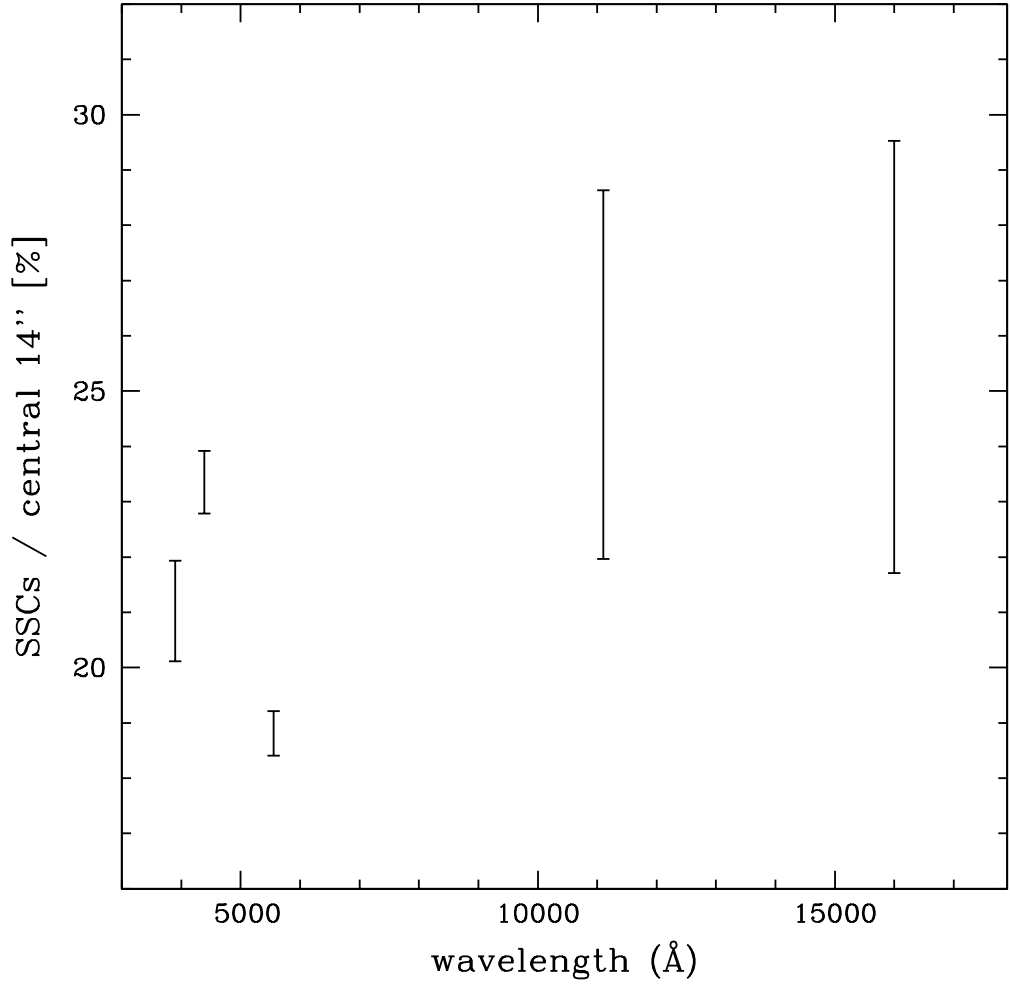}
\caption[fig]{\label{int} The percentage contribution of the SSCs to
the integrated luminosity over the central 14$''$ in different
photometric bands.
The lower and upper limits refer to zero and  average background
subtraction, respectively.
}
\end{figure}

The four young SSCs are very luminous and can provide a significant
fraction of the total light in the central region of NGC~1569.
In order to estimate such a contribution at different wavelengths,
we compare the luminosity of the clusters  to the integrated 
galaxy luminosity within a circular aperture of 14$''$ in diameter,
centered on NGC~1569-A. This region corresponds to the overlap
between the PC and NIC2 frames.
A major uncertainty in the aperture photometry computations
comes from the subtraction of the local background, particularly 
in the IR.      
As an upper limit to this background we assume the average value
measured in the less crowded regions, and as a lower limit we
assume zero background.
The relative contributions of the SSCs to the total luminosities in the
14$''$
area have been derived for both assumptions for the background.
The fractional contribution of
the four clusters in all five filters is plotted in Fig.~\ref{int}. 
The SSCs provide 20~-~25\% of the central luminosity in the blue and
near-IR, due to the dominant OB and red supergiant
populations, respectively.

A high efficiency of SF in SSCs was determined by Meurer
et al. (1995) in a large sample of starburst galaxies, and recently
confirmed, e.g., in the infrared-luminous galaxy merger NGC~3256
(Zepf et al. 1999). The fact that the efficiency is high in a wide
variety of interstellar environments (low and high metal content; low 
and high
galaxy luminosity) suggests that SF in clusters is a rather common 
process.

Aloisi et al. (2001) performed DAOPHOT PSF fitting photometry on
the NIC2 frames and found objects with $\chi^2$ and sharpness
parameters exceeding the typical values of
 well-fitted, single stars. 
Some of these anomalous stellar sources can be small star clusters which
are not properly resolved and extended in comparison
to the adopted PSF. We analyzed the properties of these cluster
candidates in an attempt to clarify their relation to the four
luminous super star clusters.
In order to isolate such possible candidate clusters from pure
blends or spurious detections we selected those objects with
F110W$\le$19.5, $\chi^2>1.5$ and sharpness~$>0.2$ in the F110W filter
and we rejected the ones in positional coincidence with the four SSCs 
for
which we obtained more accurate photometry (cf. \S~2.2).
The selected sharpness threshold is somewhat less conservative
than the 0.4 value adopted by Aloisi et al. (2001) since
here we are considering only the brightest objects.
  
\begin{figure}
%\plotfiddle{map.ps}{15cm}{0}{80}{80}{-230}{-120}
\plotone{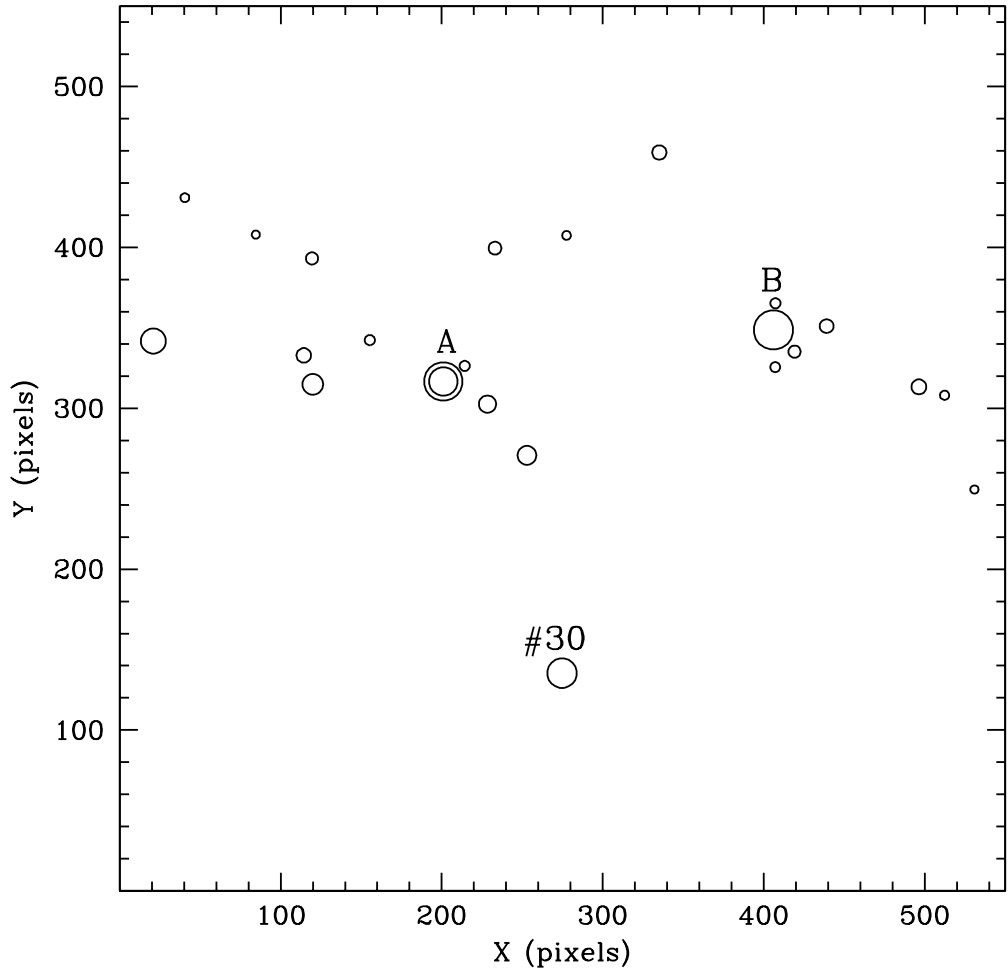}
\caption[fig]{\label{map} Spatial location of the four SSCs and
the other 20 candidate clusters in the NIC2
field of view of NGC~1569.
The spatial coordinates are in pixels (1 pixel = 0.0375$''$)    
and the symbol size is proportional to the cluster F110W luminosity.
}
\end{figure}

We identified  20 candidate clusters in our NIC2
field of view. Fig.~\ref{map} shows their spatial location. Most of
them are found
in the vicinity of the SSCs A and B. This mirrors the spatial
distribution of the clusters studied by Hunter et al. (2000), which
show a rather similar concentration towards cluster B and, even more
strongly, towards A. Not surprisingly,  
six candidate clusters of our sample show a clear counterpart
in the F555W PC field of Hunter et al. (2000; clusters
\#10, \#15, \#26, \#29, \#31, and \#32 in their Fig.~3).

\begin{figure}
%\plotfiddle{jmh.ps}{15cm}{0}{80}{80}{-230}{-120}
\plotone{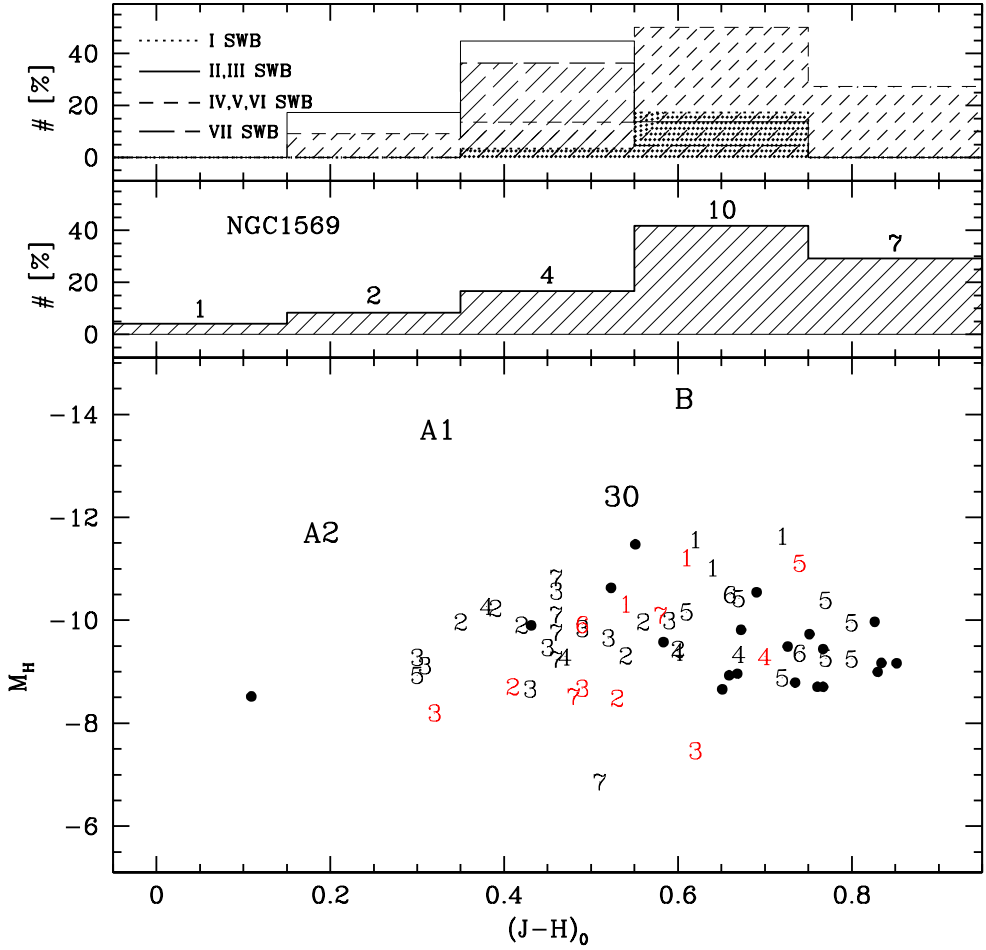}
\caption[fig]{\label{jmh}
$M_{\rm H}$ vs. $(J-H)_0$ color-magnitude diagram
(bottom) of the four SSCs (labeled A1, A2, B, 30) and
the other 20 cluster candidates (filled circles).
Numbers from 1 to 7 denote
60 Magellanic Cloud clusters, according to the age classification of
Searle, Wilkinson, \& Bagnuolo (1980; numbers 1-7 refer to the SWB types
I-VII).
Middle: histogram showing the $(J-H)_0$
color distribution of the 24 clusters
in NGC~1569. The number of objects per bin are marked.
Top: same as in the middle panel but for the
60 Magellanic Cloud clusters.}
\end{figure}

From the inferred NIC2 photometry of the four SSCs and
the other 20 cluster candidates,  and using the transformations
into the ground-based system by Origlia \& Leitherer (2000),
we  constructed the $M_{\rm H}$,
$(J-H)_0$ color-magnitude diagram
(Fig.~\ref{jmh}, bottom panel), assuming a distance modulus
$(m-M)_0=26.71$.
For comparison, we also included the integrated IR photometry
of a sample of 60 Magellanic Cloud (MC) clusters (Persson et al. 1983),
according to the age classification of
Searle, Wilkinson, \& Bagnuolo (1980, hereafter SWB). Since their
sample is infrared-selected,
even the youngest MC clusters are already in the RSG phase, hence
they have ages $>$5~Myr.

As discussed by many authors (e.g., Chiosi et al. 1986;
Girardi \& Bica 1993; Santos \& Frogel 1997; Brocato et al. 1999),
the color distributions of the MC clusters as a function of 
cluster age can be affected by systematic biases, like stochastic
effects due to the intrinsic small number of luminous red stars in
the young and intermediate age clusters, but also by some metallicity
spread among different clusters and the possible galaxy field
contamination. Bearing in mind such biases, we can
nevertheless follow a statistical approach
to investigate their color distribution and ages.    
The youngest MC clusters ($<$40~Myr, SWB type~I) have $(J-H)_0$
 between  0.4 and 0.8, with a peak distribution around 0.6~-~0.7
(see Fig.~\ref{jmh}, top panel).
The SSCs B and \#30, whose  near-IR luminosity is dominated by
the red supergiants, have indeed such colors.
The SSC A1 is at the border line, while A2 is definitively too blue
to fit into this category. This classification agrees quite well
with the age sequence we established before for the four clusters.
The MC clusters with ages between 40 and 100~Myr (SWB type II,III) have
on average slightly bluer $(J-H)_0$, with a peak around 0.4~-~0.5,
since less massive red stars contribute to their IR luminosity,
hence some contamination by a warmer component can be possible. 
Older MC clusters (SWB type IV, V, VI) with ages between a few hundred 
Myr
and a few Gyr show a peak around $(J-H)_0=0.6-0.8$. The spread is
much larger because  stochastic effects become more important in this
age range.
Finally, the oldest (and possibly most metal-poor; 
Sagar \& Pandey 1989) MC clusters of SWB type VII have $(J-H)_0\approx 
0.5$.

The majority of the 20 candidate clusters identified in our NIC2
field of NGC~1569 have $(J-H)_0$ colors consistent with SWB type~I MC
clusters (i.e., they are  RSG dominated), or with much older ones 
(SWB type IV,V,VI).
Clusters with ages around 100~Myr seem to be deficient (middle
panel of Fig.~\ref{jmh}). 
Furthermore, we cannot exclude that some of the faintest objects with 
very red
$(J-H)_0$ colors are massive AGB stars with somewhat contaminated
luminosity profiles.

\section{Conclusions}

FOS UV spectroscopy and NICMOS IR photometry of the most luminous
super star clusters in NGC~1569 allow a significant improvement over
previous relative age-dating attempts.

The four most luminous clusters
A2, A1, B, and \#30 form an age sequence from $\sim$5 to 30~Myr. A1
and A2 are the two components of the double-cluster A.
Most of the total light of A comes from A1,
with the fractional contribution of A2 increasing towards shorter
wavelengths. The subcomponents are shown
to be in slightly different evolutionary stages.
A1 is the older of the two clusters and has a significant RSG population. 
The UV spectrum of A1 and A2 combined reveals an O-star population with an
age of less than $\sim$5~Myr. The spectroscopy does not provide clues
as to whether the O stars are confined in one of the two components.
However, if NGC~1569-A behaves like a typical high-mass star-forming region,
the O population is anti-correlated with the RSGs, and is therefore
located in cluster A2. The archetypal giant H~II region 30~Doradus with
its central cluster R136 can serve as a well studied nearby test case.
Both hot and cool massive stars are found in 30~Doradus
but they are separated on
scales of a few pc. No RSGs are observed at the formation site of the
youngest O- and WR stars, and vice versa, such hot stars are
absent in Hodge~301, the cluster containing RSGs (Walborn 1984). If
R136 were observed at a distance corresponding to NGC~1569, both the
O-star and the RSG population would coincide spatially.

While the FOS spectroscopy lacks the spatial resolution for a
definitive answer, HST WFPC2 He~II imagery by Buckalew et al. (2000)
clearly demonstrated an anti-correlation of the light from RSGs and
WR stars. This is consistent with observations in the Magellanic
Clouds: WR stars of type WNL and O stars are
always observed together in metal-poor clusters and OB associations
(Massey et al. 1995), but not in the vicinity of RSGs. It is therefore
highly plausible that the O-star population in NGC~1569-A coincides
with the WR population, which in turn is anti-coincident with
the RSGs. This then implies that cluster A1 is older and contains RSGs,
and A2 is younger and hosts the O- and WR population. 
NGC~1569-B and cluster \#30 are older and fully dominated by red
supergiants, as indicated by their red IR colors.
The former is at the peak of
the RSG phase ($\sim$10 Myr), while the latter, with redder optical
colors, is slightly older, close to the end of this
evolutionary stage ($\ge$30 Myr).

The four SSCs also provide a significant fraction ($\sim$25\%) of the total
UV and IR light in the central region of the galaxy, suggesting that
highly efficient star formation per unit area occurs in
stellar clusters.

From a comparison of the optical to IR mass-to-light ratios in
NGC~1569-A1
with evolutionary synthesis models,
we find that a standard Salpeter IMF with lower mass cutoff
down to the hydrogen-burning limit
is consistent with the data. Unless gravitational forces lead to 
cluster disruption,
NGC~1569-A1 can evolve into a system whose properties are quite
similar to those of  present-day globular clusters.

Other candidate stellar clusters, less luminous than the four SSCs
discussed above, have been identified in the central
region of NGC~1569.
They span a  wide range of ages,
even though the majority of them have near IR colors
more consistent with a dominant, young stellar population
in the RSG phase. In that respect the clusters in
NGC~1569 resemble the populous young clusters in the Magellanic
Clouds.

\acknowledgments

We thank G. De Marchi for providing the optical data-set and
F. R. Ferraro for his support and suggestions during 
the photometric reduction. This work has been partly supported by the
Italian ASI, through grant ARS-99-44, and Murst, through Cofin2000.
Support for this work was provided
by NASA through grant number GO-06111.01-94A from the Space Telescope
Science Institute, which is operated by the Association of Universities
for Research in Astronomy, Inc., under NASA contract NAS5-26555.

\newpage

\begin{deluxetable}{lcccccc}
\rotate
\tablewidth{18.8truecm}
\tablecaption{\label{colors} De-reddened magnitudes and colors 
colors for the four SSCs in NGC~1569.}
\tablehead{
\colhead{SSC} &
\colhead{(555W)$_0$} &
\colhead{(380W-439W)$_0$} &
\colhead{(439W-555W)$_0$} &
\colhead{(110W-160W)$_0$} &
\colhead{(555W-110W)$_0$} &
\colhead{(439W-160W)$_0$}
}
\startdata
A1  &  13.7 & -0.5 &  +0.0  &   0.5 &  +0.2 & +0.7\\
A2  &  15.1 & -0.5 &  -0.1  &   0.3 &  -0.2 & +0.0\\
B   &  14.5 & -0.5 &  +0.3  &   0.9 &  +1.2 & +2.4\\
\#30   &  16.3 & -0.1 &  +0.4  &   0.8 &  +1.2 & +2.4\\
\enddata
\end{deluxetable}

\clearpage

\clearpage

\clearpage

\clearpage

\clearpage

\clearpage

\clearpage

\clearpage

\clearpage

\end{document}